\documentclass[12pt]{iopart}

\usepackage{amssymb}
\usepackage{graphicx}
\usepackage{bm}
\usepackage{epsfig}
\usepackage{amsfonts}

\begin{document}

\title{Impurity Effects in Two-Electron Coupled Quantum Dots: Entanglement Modulation}

\author{Diego S. Acosta Coden, Rodolfo H. Romero, Alejandro Ferr\'on
 and Sergio S. Gomez}

\address{ Instituto de Modelado e Innovaci\'on Tecnol\'ogica (CONICET-UNNE) \\ and Facultad de Ciencias Exactas y Naturales y Agrimensura, Universidad Nacional del Nordeste,
Avenida Libertad 5400, W3404AAS Corrientes,
Argentina}

\begin{abstract}

We present a detailed analysis of the electronic and optical properties of
two-electron quantum dots with a two-dimensional Gaussian confinement 
potential. 
We study the effects of Coulomb impurities and the possibility 
of manipulating the entanglement of the electrons by controlling the
confinement potential parameters. 
The degree of entanglement becomes highly modulated by both the location 
and charge screening of the impurity atom, resulting in two regimes: 
one of low entanglement and other of high entanglement, with both of them 
mainly determined by the magnitude of the charge. 
It is shown that the magnitude of the oscillator strength of the system could provide 
an indication of the presence and characteristics of impurities and, therefore, the degree of entanglement.

\end{abstract}
\date{\today}

\maketitle

\section{Introduction}

Semiconductor quantum dots (QDs) are excellent candidates for realizing
qubits for quantum information processing because of the potential for 
excellent manipulability and scalability. In contrast
to real atoms and molecules, in QDs the electronic and optical
properties are highly tunable. Tremendous advances in semiconductor technology
allow the preparation of complex structures and give the possibility to the 
experimentalists to have a great control on the parameters that define
the electrical and optical properties of these systems \cite{QD1}. 

It is known that the presence of impurity centers has a great influence on 
the optical and electronic properties of nanostructured
materials. Since the pioneering work of Bastard \cite{bast} many authors
have investigated the effects of impurities on different properties
of artificial atoms and molecules. A recent work \cite{dasarma1} 
studied the effects of having unintentional charged impurities 
in two-electron laterally coupled two-dimensional double quantum-dot 
systems. They analyzed the effects of quenched random-charged impurities 
on the singlet-triplet exchange coupling
 in two-electron double quantum-dots. Although
there is an enormous interest in applying these systems in quantum
information technologies, there are few works trying to quantify 
the effect of charged impurities on this kind of tasks. 
The existence of unintentional impurities, which are always present
in nanostructured devices, affects seriously the possibility of using these 
devices as quantum bits. Although the distribution
and concentration of impurities in these systems result unknown
parameters, there are some recent works that propose the possibility
of experimentally control these issues \cite{ex1,ex2,ex3}. Impurity 
doping in semiconductor materials is considered as a useful
technology that has been exploited to control optical and
electronic properties in different nanodevices. 

It is worth to mention that, due to environmental
perturbations, these systems lose coherence. For example, confined electrons 
interact with spin nuclei through the hyperfine interaction leading, 
inevitably, to decoherence \cite{PettaDec}. Even, having 
just one  charged impurity could induce qubit decoherence if this impurity is 
dynamic and has a fluctuation time scale comparable to gate operation time 
scales \cite{dasarma1}. Decoherence is a phenomenon that plays a central 
role in quantum information and its technological applications 
\cite{dec1,dec2,dec3,dec4,dec5,dec6,dec7}.

The entanglement, which is one of the most curious phenomena in quantum
mechanics, is being considered in recent years as a physical resource that
can be used for quantum information processes as teleportation of quantum 
states \cite{tichi,amico,damico1}. There exists the possibility of
manipulate the amount of entanglement in a QD molecule by controlling
the nanostructure parameters that define the nanodevice. Zunger and He
 \cite{zu1} studied the effect of interdot distance and asymmetry on 
the spatial entanglement of two-electron coupled quantum dots. They showed 
that the asymmetry in these systems
significantly lowers the degree of entanglement of the two electrons.
Two-electron entanglement of different quantum dots atoms and molecules
have been studied by several authors in the last decade \cite{damico1,zu1,zu2,
damico2,damico3,os1,os2,fos1,pots,pos,ana}. The presence of nearby charged impurities, as Das Sarma and Nguyen show,
have an important effect on the singlet-triplet coupling, with unwanted impact in  
quantum information tasks. One of the main goals of this paper is 
the calculation of the spatial entanglement \cite{damico1,damico2,os1} 
of the two electrons in a double QD molecule in presence of 
charged impurities. Experimentally, results very difficult to measure
the amount of entanglement of two electron in a coupled QD directly.
There exist several techniques that allows one to measure the possibility
of double occupation \cite{zu1} and the optical properties such as the 
dipole transition, the oscillator strength and
the photoionization cross section \cite{opt1,opt2,opt3,opt4,opt5,opt6}  
of these systems.
If we know the relationship between these quantities and the degree of
entanglement, we can have information about the amount of entanglement
of our system. The possibility of using this information in order 
to design nanodevices according to the level of entanglement desired results
quite difficult because the positions and the strength of the impurities
are unknown. Despite this, there are some recent experiments which show 
the mechanism of dopant incorporation and how the incorporation 
of impurity defects can be controlled \cite{ex1,ex2,ex4}.

The aim of this work is to present a detailed analysis of the electronic 
and optical
properties of a two-dimensional two-electron coupled quantum dot and the
effect of impurities. In particular we show that the entanglement of 
the electrons is strongly modulated by the position and charge of the 
impurity. We also show that optical measurements would allow to 
obtain information about the effect of the impurity in these kind of devices.
The paper is organized as follows. 
In Section II, we introduce the model for the two-dimensional two-electron coupled quantum dot and briefly describe the method used to calculate its electronic structure.
In Section III, we calculate the spatial entanglement in the presence of one impurity and discuss its relation to the exchange coupling. 
Section IV, contains calculations of the oscillator strength, for a range of parameters of the system, that show the modifications of the 
optical properties in the presence of a charged impurity, with the aim of allowing to correlate optical measurements with the degree of entanglement of the system. 
Finally, In Section V we summarize the conclusions with a discussion of 
the most relevant points of our analysis.

\section{Model and calculation method}

We consider two laterally coupled two-dimensional quantum dots whose 
centers are separated a distance $d$ from each other, and containing two 
electrons. In quantum dots electrostatically produced, both their size and 
separation can be controlled by variable gate voltages through metallic 
electrodes deposited on the heterostructure interface. The eventual existence 
of doping hydrogenic impurities, probably arising from Si dopant atoms in the 
GaAs quantum well, have been experimentally studied \cite{Ashoori92}. These 
impurities have been theoretically analyzed with a superimposed attractive 
$1/ r$-type potential \cite{im1,im2}.
Furthermore, some avoided crossing and lifted degeneracies in the spectra 
of single-electron transport experiments have been attributed to negatively 
charged Coulomb impurities located near to the QD \cite{ex5}. From fitting 
the experimental transport spectra to a single-electron model of softened 
parabolic confinement with a Coulomb charge $q$, a set of parameters are 
obtained; among them, a radius of confinement of $15.5$ nm, a confinement 
frequency $\hbar\omega=13.8$ meV and an impurity charge of approximately 
1 or 2 electron charges. Indeed, the 
uncertainty in the parameters and the suppositions introduced in the model 
does not allow one to precisely ensure the impurity charge, with the 
screening probably reducing its effective value to less than an electron 
charge. Therefore, we consider the charge of the doping atom $Ze$ as a 
parameter varying in the range $0\leq Z \leq 1$, in order to explore its 
effect on the properties of the system.

In this work we model the Hamiltonian of the two-dimensional two-electron 
coupled quantum dot in presence of charged impurities within the single 
conduction-band effective-mass approximation \cite{mod1}, namely,

\begin{equation}
\label{h1}
H=h({\bf r}_1)+h({\bf r}_2)+\frac{e^2}{4\pi\varepsilon \varepsilon_0 r_{12}},
\label{Hamiltonian}
\end{equation}

\noindent where ${\bf r}_i=(x_i,y_i)$ ($i=1,2$) and

\begin{equation}
\label{h2}
h({\bf r})=-\frac{\hbar^2}{2m^*}\nabla^2+V_L({\bf r})+V_R({\bf r})+V_A({\bf r}),
\end{equation}

\noindent where $h({\bf r})$ is the single-electron Hamiltonian that includes 
the kinetic energy of the electrons, in terms of their effective mass 
$m^*$, and the confining potential for the left and right quantum dots $V_L$ 
and $V_R$, and the interaction of the electrons with the charged impurities, 
$V_A$. 

The last term of the Hamiltonian, Eq. (\ref{Hamiltonian}), represents the 
Coulomb repulsive interaction between both electrons at a distance 
$r_{12}=|{\bf r}_2-{\bf r}_1|$ apart from each other, within a material of 
effective dielectric constant $\varepsilon$.
We model the confinement with Gaussian attractive potentials

\begin{equation}\label{h3}
V_{i}({\bf r})=-V_0\exp\left(-\frac{1}{2a^2} |{\bf r}-{\bf R}_{i}|^2\right), \ (i=L, R),
\end{equation}

\noindent where ${\bf R}_L$ and ${\bf R}_R$ are the positions of the center 
of the left and right dots, $V_0$ denotes the depth of the potential and $a$ 
can be taken as a measure of its range.
Along this work, we will consider a single impurity atom centered 
at ${\bf R}_A$, and modelled as a hydrogenic two-dimensional Coulomb potential

\begin{equation}\label{h4}
V_A({\bf r})= -\frac{Ze^2}{4\pi\varepsilon \varepsilon_0 
|{\bf R}_A-{\bf r}|}
\end{equation}

Since the Hamiltonian does not depend on the electron spin, its eigenstates 
can be factored out as a product of a spatial and a spin part

\begin{equation}
\Psi_i({\bf r}_1,{\bf r}_2,m_{s_1},m_{s_2}) = \Psi_i^{S}({\bf r}_1,{\bf r}_2) \chi_{S,M},
\end{equation}

\noindent where $S=0$, 1 for singlet and triplet states, respectively, 
and $M = m_{s_1} + m_{s_2}$ is the total spin projection.

The eigenstates of the model Hamiltonian can be obtained by direct 
diagonalization in a finite basis set \cite{var1}. 
The spatial part is obtained, in a full configuration 
interaction (CI) calculation, as 

\begin{equation}\label{variational-functions}
\Psi_m^S({\bf r}_1,{\bf r}_2) = \sum_{n=1}^{N_{\rm conf}} c^S_{mn} \Phi^S_{n}({\bf r}_1,{\bf r}_2)
\end{equation}

\noindent where $N_{\rm conf}$ is the number of singlet ($S=0$) or triplet 
($S=1$) two-electron configurations $\Phi^S_{n}({\bf r}_1,{\bf r}_2)$ 
considered, and $n=(i,j)$ is a configuration label obtained from the indices 
$i$ and $j$ from a single electron basis, i.e.,

\begin{equation}
\Phi^S_{n}({\bf r}_1,{\bf r}_2) = \frac{1}{\sqrt{2}}\left[ \phi_i({\bf r}_1)\phi_j({\bf r}_2)+(1-2S)\phi_j({\bf r}_1)\phi_i({\bf r}_2)\right]
\end{equation}

\noindent for $i\ne j$, and $\Phi^{S=0}_{n}({\bf r}_1,{\bf r}_2)=
\phi_i({\bf r}_1)\phi_i({\bf r}_2)$ for the doubly occupied singlet states.

We chose a single-particle basis of Gaussian functions, centered at the dots 
and atom positions ${\bf R}_P$ ($P = L, R, A$), of the type \cite{sr1,sr2}

\begin{equation}\label{variational-functions}
\phi_{i}({\bf r})=N x^{m_i} y^{n_i} \exp\left(-\alpha_i|{\bf r}-{\bf R}_P|^2\right), 
\end{equation}

\noindent where $N$ is a normalization constant, and $\ell_i=m_i+n_i$ 
is the $z$-projection of the angular momentum of the basis function. 
The exponents $\alpha_i$ were optimized for a single Gaussian well and 
a single atom separately, and supplemented with extra functions when used 
together.
For our calculations a basis set of $2s 2p$ functions for the dots, 
and $5s 5p 1d 1f$ for the atom was found to achieve converged results for 
the energy spectrum.  

The numerical results presented in this work refers to those corresponding 
to the parameters of GaAs: effective mass $m^*=0.067 m_e$, effective 
dielectric constant $\varepsilon=13.1$, Bohr radius $a^*_B=10$ nm and 
effective atomic unit of energy 1 Hartree$^*=10.6$ meV 
\cite{ex5,dasarma1}. The depth of the Gaussian potentials modelling the dots 
are taken as $V_0=4$ Hartree$^*=40.24$ meV, and its typical range 
$a=\sqrt{2}a_B^*=14.1$ nm.

\section{Entanglement entropy and exchange coupling}

The proposed applications of QDs for quantum computing require a large 
exchange coupling between electrons along separated regions of space. To some 
extent, both requirements compete with each other. In a simple picture, one 
could have a large exchange coupling for electrons doubly occupying the same 
dot or atom. In such a case, the singlet state has the form of a product 
wave function $\varphi_0({\bf r}_1) \varphi_0({\bf r}_2)$ with the 
corresponding singlet spin function; the lowest triplet state, however, 
has the form of the antisymmetrized product of two single-electron functions, 
$\varphi_0({\bf r}_1) \varphi_1({\bf r}_2)-\varphi_0({\bf r}_2) 
\varphi_1({\bf r}_1)$, of different single-particle energies $\varepsilon_0$ 
and $\varepsilon_1$. Thus, the triplet state will have a quite higher energy 
than the singlet state, thus giving a large exchange coupling. Nevertheless, 
such a large coupling is not favourable for quantum computing tasks because the 
states are localized in space. Using electron states as qubits requires, 
for instance, the feasibility to detect the single or double occupancy of 
two quantum dots, separated a measurable distance, while keeping both 
electrons correlated.
 
\begin{figure}[ht]
\begin{center}
\includegraphics[width=6cm]{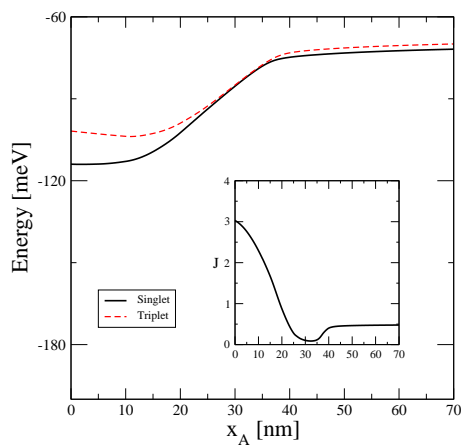}
\end{center}
\caption{\label{f1} (Color online) Calculated ground-state energy of 
the two-electron double QD, for an interdot distance $d=30$ nm, with one 
impurity 
atom of charge $Z=1$ located along the interdot axis as a function of impurity 
position. The black line (red-dashed line) shows the singlet (triplet) 
ground-state energy. The inset shows the singlet-triplet exchange coupling $J$  
as a function of impurity position.}
\end{figure}

As the interdot separation increases, the electron-electron interaction 
diminishes and its relative importance with respect to the confining 
potential tends to vanish. In the limit of large interdot separations, 
the Coulomb repulsion is minimized by singly occupying each quantum dots 
with an electron.
In such a limit, the energies of both the singlet $(+)$ and triplet $(-)$ 
states,  $\varphi_0({\bf r}_1) \varphi_1({\bf r}_2) \pm \varphi_0({\bf r}_2) 
\varphi_1({\bf r}_1)$ approach the sum of singly occupied dots and their 
difference $J$ tends to zero.
In other words, the best conditions for applications to quantum information 
processing arises from a compromise between a high spatial correlation of 
pairs of electrons at the longest possible lengths where the exchange 
coupling $J$ is still sizable. This behaviour is illustrated in Fig. 
\ref{f1}, assuming a positively charged impurity of one electron charge 
($Z=1$).

Fig \ref{f1} shows the singlet and triplet ground-state energies for the 
double QD, separated a distance $d=30$ nm, as a function of impurity position 
$x_A$. The inset shows the behaviour of the singlet-triplet exchange coupling
 as 
a function of the impurity position. These results are in qualitative 
agreement with those of Ref. \cite{dasarma1}. The singlet-triplet exchange 
coupling 
is strongly affected for the presence of the charged impurity, it has the 
maximum value when the impurity is centered in between the two dots, 
and it has a minimum close to zero when the impurity is located at 
$x_A=d$. Expectedly, the splitting goes asymptotically to the impurity-free 
double QD case when the impurity atom is located far away from the double 
QD system. 
We shall show below that the impurity positions that give high energy 
splitting, i.e., those near to the middle of the interdot distance, 
correspond to a two-electron ground state whose spatial wave function is 
highly localized at the impurity atom, thus having a small spatial entanglement.

In what follows we shall restrict ourselves to the impurity located 
along the interdot $x$-axis, ${\bf R}_A=(x_A,0)$. 

We shall study now how the degree of spatial quantum correlation of two 
electrons in the coupled QD is modified by the position and charge of a 
screened 
atomic impurity.
As mentioned above, the eigenstate wave functions can be factorized in its 
orbital and spin part. For the ground state, the spin part is a singlet wave 
function, which is maximally entangled and constant. Therefore, throughout 
this work, we will only consider the spatial entanglement 
\cite{damico1,damico2,os1,os2,fos1}.

The Von Neumann entropy of the reduced density matrix quantifies the 
entanglement for a bipartite pure state and can be calculated using
\cite{damico2,os1,os2,fos1,pots}

\begin{equation}\label{von-neumann-entropy}
S = -\mathrm{Tr}(\hat{\rho}^{\mathrm{red}}
\log_2{\hat{\rho}^{\mathrm{red}}}) ,
\end{equation}

\noindent where  $\hat{\rho}^{\rm red}={\rm Tr}_2|\Psi\rangle\langle\Psi|$ is 
the reduced density operator, $\Psi$ is the two-electron wave function and
the trace is taken over one electron. The Von Neumann entropy could be 
written as

\begin{equation}\label{von-neumann-entropy2}
S = -\sum_i \lambda_i \log_2 \lambda_i ,
\end{equation}

\noindent where $\lambda_i$ are the eigenvalues of the spatial part of the
reduced density operator
\begin{equation}
\int \rho^{\mathrm{red}}({\bf r}_1,{\bf r}^{\prime}_1) \phi_i({\bf r}^
{\prime}_1)\; d{\bf r}^{\prime}_1 = \lambda_i \phi_i({\bf r}_1) \, .
\end{equation}

\noindent where

\begin{equation}
\rho^{\mathrm{red}}({\bf r}_1,{\bf r}_1^{\prime}) = \int
\Psi^{\ast}({\bf r}_1,{\bf r}_2) \Psi({\bf r}_1^{\prime},
{\bf r}_2) \; d{\bf r}_2.
\end{equation}

\begin{figure}[ht]
\begin{center}
\includegraphics[width=6cm]{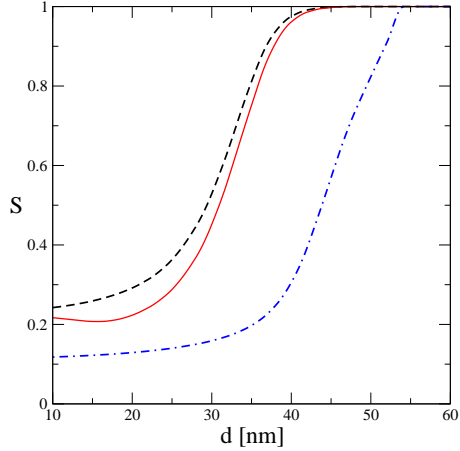}
\end{center}
\caption{\label{sic}(Color on-line) Von Neumann entropy of the reduced
density matrix for the two-electron coupled QD as a function of the
interdot distance. Black-dashed line shows the entropy when there is no
impurity present in the sample, the red line corresponds to the
entropy when a single impurity $Z=0.1$ is located at $x=y=0$ and blue-dotted 
dashed line corresponds to $Z=1.0$.}
\end{figure}

\noindent Fig. \ref{sic} shows the Von Neumann entropy for two electrons 
in the double QD as a function of its interdot distance, in absence of 
impurity (black-dashed line) and with atomic charges  $Z=1$ 
(blue-dashed doted line) and $Z=0.1$ (red solid line) located at the 
center of the double QD. 
In all the cases it is observed that, for small interdot separations, 
the entropy is small, smoothly increasing with the interdot separation. 
The increase of the spatial entanglement is due to a gradual delocalization 
of the ground state wave function. 
For interdot distances between 20 and 40 nm, there is a large increase of 
the entanglement entropy, signalling a qualitative change of the ground 
state wave function from a atomic doubly occupied state to a state with 
both dots singly occupied, reached at large interdot separations 
($d\gtrsim 50$ nm), where the entropy saturates to its maximum $S=1$.
The variation of $S$ is similar for all the cases, although the presence of 
the impurity decreases the entanglement for every interdot separation, 
due to the fact that the atomic potential contributes to localize the 
electronic density at the center of the system.

\begin{figure}[ht]
\begin{center}
\includegraphics[width=8cm]{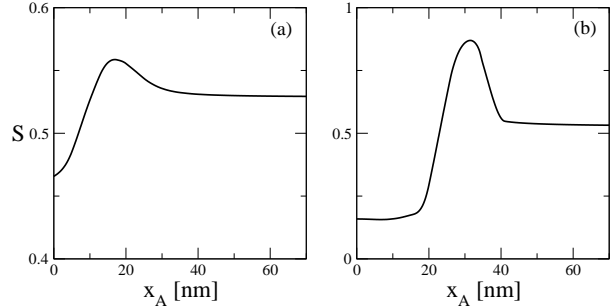}
\end{center}
\caption{\label{svx2z} Von Neumann entropy of the reduced
density matrix for the two-electron coupled QD as a function of the
impurity position (along the interdot axis) for $d=30\, nm$ and different 
impurities strength: (a) $Z=0.1$ and (b) $Z=1.0$.}
\end{figure}

\noindent The effect of the charge and location of the impurity on the 
spatial entanglement, for fixed QDs geometry, can be observed at Fig. 
\ref{svx2z} where the entropy is depicted as a function of the impurity 
position. The separation between the two QDs is kept  fixed at 30 nm, and 
two limiting cases are considered: a highly screened atomic charge $Z=0.1$ 
and a unscreened charge $Z=1$.
In both cases, the entanglement entropy increases as the impurity moves 
off the center of the double QD until a position where $S$ reaches a 
maximum, finally decreasing to a value $S=0.53$, when the atom is distant 
from the dots ($x\gtrsim 40$ nm). 

\noindent The minimum and maximum of entanglement produced by the small 
charge $Z=0.1$ are less pronounced than those due to the highly charged 
impurity $Z=1$. 
This modulation of the entanglement by the impurity position reflects the 
existence of two regimes: one of low entanglement for impurities at 
(or near to) the center of the interdot distance, and another of higher 
(but not maximum) entanglement for atomic positions external to the interdot 
segment. In Fig. \ref{svx2z}, this two regions are the ones to the left and the 
right of the bell-shaped peak of $S$, respectively. The peak position itself 
depends on the magnitude of the charge. For small charges, the maximum degree 
of entanglement occurs at $x_A\approx 17$ nm, that is, close to the center 
of the dot to the right. For the large charge $Z=1$, however, the peak of $S$ 
occurs at $x\approx 30$ nm. The rationale for it is that, for low charged 
impurities, the atomic potential is a weak perturbation to the QDs potential 
wells. Therefore, the entropy varies in a small range ($0.47\leq S \leq 0.56$) 
around the impurity-free case $S=0.53$. For large impurity charges, 
nevertheless, the atomic potential is strong and the position of its center 
greatly determines the spatial wave function. The range of atomic positions 
($0\leq x_A \lesssim 20$ nm) along which $S$ remains low, can be understood 
as due to the localization of the electrons close to the atom. When the atom 
is inside one of the dots (QD$_R$), the atomic potential reinforces that 
of the dot well and the electron density localization, thus giving a low 
degree of entanglement. When the atom moves towards outside the double QD, 
the strength of the double well competes with the large atomic potential until 
an atom-double QD distance of {\em ca.} 30 nm, where becomes energetically 
convenient to delocalize the electron wave function, resembling the double QD 
bond in absence of impurity.

\begin{figure}[ht]
\begin{center}
\includegraphics[width=8cm]{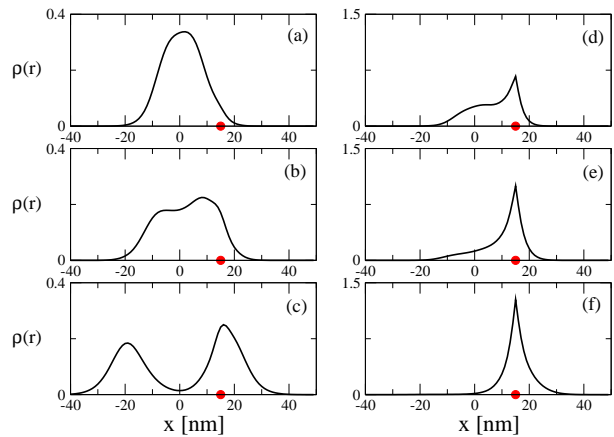}
\end{center}
\caption{\label{roc}(colour on-line) Ground state one-electron density along
the interdot axis. Left panels show the weak impurity limit $Z=0.2$ and 
right panels the strong impurity limit $Z=0.8$:
(a) and (d) $d=15 nm$, (b) and (e) $d=25 nm$ and (c) and (f) $d=40 nm$
. Red circles show the impurity position $x_A=15 nm$ }
\end{figure}

To show clearly the influence of the atomic charge on the wave function, 
let us consider two a bit less extreme situations: $Z=0.2$ and $Z=0.8$. 
Fig. \ref{roc} shows the ground state electron density along the interdot 
axis when the impurity atom is located at $x_A=15$ nm, for three different 
interdot separations, $d=15$, 25 and 40 nm. The panels to the left show 
that for the small charge, as the QDs separate from one another, the electron 
density develops peaks located at the potential well centers. For the large 
charge $Z=0.8$, however, the density is always peaked at the impurity position. 
Therefore, in this last case, the presence of the impurity could spoil the 
performance of the device for quantum computing tasks due to the high 
localization of the electron density entails a low degree of entanglement.

\begin{figure}[ht]
\begin{center}
\includegraphics[width=8cm]{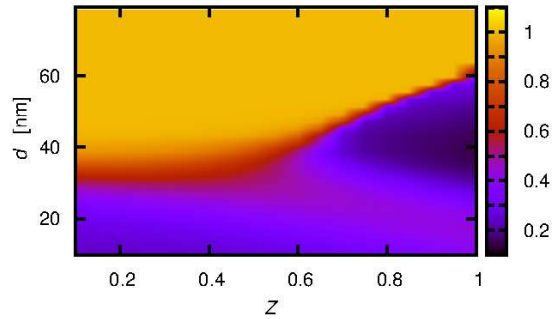}
\end{center}
\caption{\label{smap}(colour on-line) Contour map for the Von Neumann entropy 
of the reduced density matrix for the two-electron coupled QD as a 
function of impurity strength and interdot distance for $x_A=15$ nm.}
\end{figure}

\noindent Fig. \ref{smap} shows the dependence of the Von Neumann entropy 
on the impurity charge and interdot distance for a given position of the 
impurity atom: $x_A=15$ nm.
Ideally, providing that the value of the impurity charge could be measured in 
a given sample, one would be able to choose the optimal interdot distance 
for a given degree of entanglement. The figure clearly shows the 
aforementioned regimes of weak ($Z\lesssim 0.6$) and strong ($Z\gtrsim 0.6$) 
impurity potential, corresponding to low and high degree of entanglement, 
respectively.
For a given (fixed) small impurity charge $Z$, the entropy increases 
monotonically as the interdot distance $d$ increases. On the other hand, 
for a given large $Z$, by increasing the distance $d$, the entropy increases 
for small distances $d$ up to a maximum value, diminishes to a minimum and 
sharply increases again until its asymptotic impurity-free value $S=1$.

\begin{figure}[ht]
\begin{center}
\includegraphics[width=8cm]{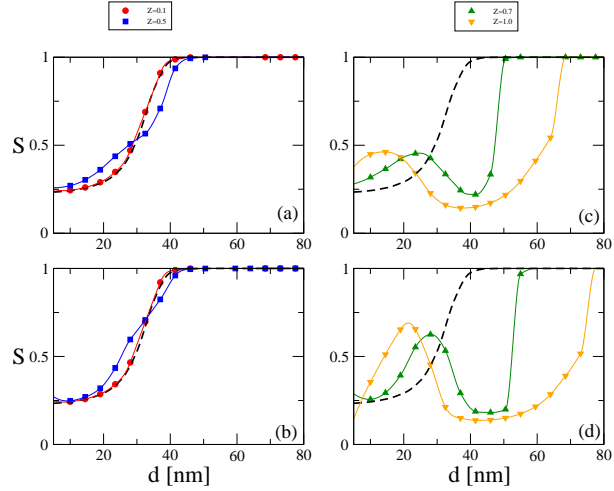}
\end{center}
\caption{\label{svdc}(colour on-line) Von Neumann entropy of the reduced
density matrix for the two-electron coupled QD as a function of the
interdot distance for $x_A=15\, nm$ ((a) and (c)), $x=20\, nm$ ((b) and (d)) 
and different values
of the impurity Strength. (a) and (b) show the weak impurity limit ($Z=0.1$ and
$Z=0.5$) while in (c) and (d) we observe the strong impurity limit
($Z=0.7$ and $Z=1.0$). The black-dashed line shows the behavior of
the entropy when there is no impurities in the sample. }
\end{figure}

\noindent Fig. \ref{svdc} shows the dependence of the entropy on the interdot 
distance, for different impurity positions ($x_A=15$ and 20 nm) and charges 
($Z=0.1$, 0.5, 0.7 and 1). The corresponding variation in absence of impurity 
is also represented in dashed lines for reference.
It can be seen two qualitatively distinct behaviours for small ($Z=0.1$, 0.5) 
and large ($Z=0.7$, 1) charges . The monotonic increase of $S$ with $d$ 
is characteristic of the weak atomic potential; separating the QDs with a 
small atomic charge in between of them, produces little changes in the 
electron distribution as compared with the impurity-free double QD. On the 
other hand, strong atomic potentials induce a modulation of the entropy as 
$d$ increases; for small values of $d$, all three potentials are close to 
each other and the electron density localizes around their centers. For large 
interdot distances, the energy of the system is minimized by decreasing the 
electronic repulsion, i.e., by delocalizing the wave function and, hence, 
increasing its entanglement. 

\begin{figure}[ht]
\begin{center}
\includegraphics[width=8cm]{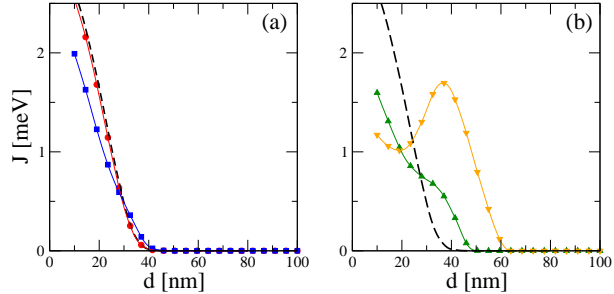}
\end{center}
\caption{\label{fj}(colour on-line) J as a function of the interdot
distance $d$ with an impurity center located in $x_A=15 nm$. On the left panel
(a) we observe the singlet-triplet coupling for $Z=0.1$ (red circles) and
$Z=0.5$ (blue squares). In (b) we show  the singlet-triplet coupling for
$Z=0.7$ (green up triangles) and $Z=1.0$ (orange down triangles). The
black-dashed line represents the singlet-triplet with no impurity.}
\end{figure}

\noindent Fig. \ref{fj} shows this effect on the exchange coupling 
corresponding to situations of Fig. \ref{svdc}(a) and \ref{svdc}(c), having 
the atom at $x_A=15$ nm. It can be observed that $S$ and $J$ have, roughly, 
opposite variations; whence the atomic potential is weak, $S$ increases 
and $J$ decreases as the QDs separate from each other. In the regime when 
the atomic potential is strong, the maximum of $S$ occurs at the minimum of 
$J$ and reciprocally; furthermore, at large QDs separations, as the entropy 
goes to its asymptotical value $S=1$ the exchange coupling tends to zero. 
Then, for specific quantum information applications, it could be desirable 
to tune the interdot distance for harnessing one or both properties.

The variety of behaviours of the degree of spatial entanglement with the 
various parameters of the system, described in this section, is rooted in 
the spatial distribution of the electron wave function. We shall discuss in 
the next section a relation with an optical property, like the oscillator 
strength, in order to provide a feasible connection with measurable 
magnitudes.

\section{Impurity effect on the optical properties}

The optical susceptibility of a system depends on its transition amplitude 
for the interaction of its dipole moment with the optical electric 
field between two singlet states $\Psi_i$ and $\Psi_j$, say the ground and 
an excited states, and the corresponding energy differences. The oscillator 
strength for an electric field applied along the interdot axis

\begin{equation}
f_{ij}=\frac{2m^\ast}{\hbar} (E_j-E_i) |\langle \Psi_0|x_1+x_2
|\Psi_1\rangle|^2, 
\label{osf}
\end{equation}

\noindent takes both magnitudes into account and provides information on the 
feasibility of such optical excitations.

\begin{figure}[th]
\includegraphics[width=8cm]{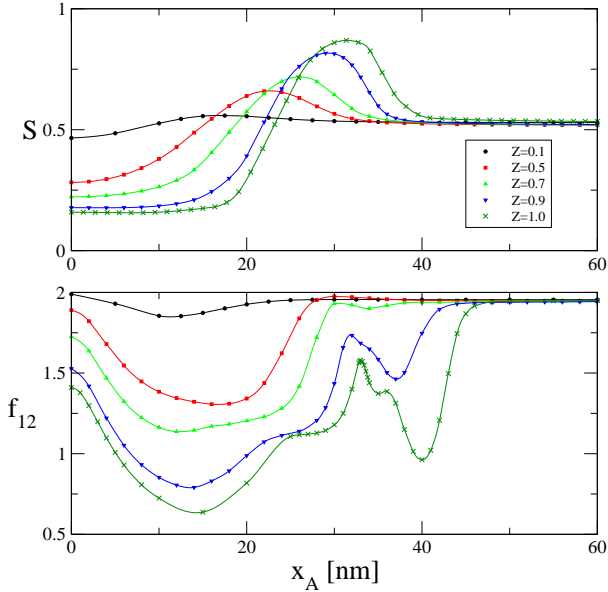}
\caption{\label{OS vs x}  Von Neumann entropy of the reduced
density matrix (upper panel) and Oscillator strength $f_{12}$ (lower panel) 
between the ground and first excited singlet state of a double quantum dot with an impurity of charge $Z$ as a function of the impurity position $x_A$, for different values of $Z$.}
\end{figure}

We study here how the impurity affects the oscillator strength of the
double QD. The dots are kept 30 nm separated from each 
other and the position of the impurity $x_A$ is varied from the center of 
interdot segment ($x_A=0$) to a large separation from the dots 
($x_A=70$ nm), including the case of the impurity centered in one dot 
($x_A=15$ nm). The charge $Z$ of the atom is varied from $Z=0.1$ 
(highly screened impurity) to $Z=1$ (low screening).
For the system considered, the oscillator strength between the ground and 
the first excited singlet states, $f_{12}$, is the dominant contribution 
with respect to all others $f_{ij}$. The precision of the calculation was 
checked by verifying the Thomas-Reiche-Kuhn sum rule, $\sum_{ij} f_{ij} = N$, 
with $N=2$ being the number of electrons in the system. The results are shown 
in Fig. \ref{OS vs x} together with the entanglement entropy for the same 
atomic positions and charges.

The cases of weak and strong electron-atom Coulomb interaction are 
clearly distinguishable. In the regime of small impurity charge 
($Z\lesssim 0.6$), the oscillator strength $f_{12}$ varies approximately 
with a quadratic dependence on $x_A$; i.e., it starts from 
$f_{12}\approx 2$, reaches a minimum around $x_A=15$ nm, to finally increases 
up to a value of 2, at nearly $x_A=30$ nm. The larger the impurity charge $Z$, 
the more pronounced the minimum of $f_{12}$. Placing the atom further away 
from the double dot system ($x_A>30$ nm) does not change $f_{12}$. 

On the other hand, in the regime of high impurity charge ($Z\gtrsim 0.6$), 
the oscillator strength $f_{12}$ exhibits richer features as compared to the 
small charge case. The most remarkable behavior correspond to $Z=1$ which 
successively shows a similar decreasing, from $f_{12}=1.4$ at $x_A=0$, to 
$f_{12}=0.6$ at $x_A=15$ nm, followed by an increase up to $x_A=25$ nm, a 
small plateau around 30 nm, a peak at $x_A=33$ nm, a minimum of 
$f_{12}\approx 1$ at 40 nm, finally approaching the saturation value 
$f_{12}=2$ for $x_A\gtrsim 50$ nm. For intermediate $0.5\leq Z \leq 1$ values, 
a gradual transition between both regimes is observed; namely, by decreasing 
$Z$ from 1 to 0.5, the minimum of the region $x_A\approx 40$ nm becomes 
shallower, the peak is softened, and the plateau merges with the minimum 
occurring at 15 nm, thus giving the flat minimum of the weak impurity regime.

It should be noted that the range $0\leq x \leq 15$ nm corresponds to an 
impurity atom located in between the dots, while for $x_A\geq 15$ nm, the atom 
is outside the segment defined by the interdot centers. Consequently, the 
existence of an impurity into the system would cause a diminishing in $f_{12}$ 
and, therefore, in the light absorption or emission of the double dot device. 
This effect is stronger the closer is the atom to one dot. The most favourable 
situation for optical excitation (high $f_{12}$) correspond to an impurity 
centered in between the dots or outside the interdot separation, faraway from 
any of them.\\

We shall discuss in the following, the behavior of the oscillator strength 
as due to changes in the electronic structure induced by the variation of 
the position of the impurity, starting by the most striking case of a highly 
charged impurity $Z=1$. 
We displace the atom along the line joining both dots, which we take as the 
$x$-axis; therefore, we consider the two-particle wave function along the $x$ 
axis for the coordinates $x_1$ and $x_2$ of each 
electron

\begin{equation}
\Psi_i(x_1,x_2) = \Psi_i({\bf r}_1,{\bf r}_2) = \Psi_i(x_1,0;x_2,0),
\end{equation}

\noindent for the two lowest singlet states 
$\Psi_i({\bf r}_1,{\bf r}_2)$, $i=0$ (ground state) and $i=1$ 
(first excited state). The function $\Psi(x_1,x_2)$, represented as a 
two-dimensional plot in the $(x_1,x_2)$-plane, allows one to visualize the 
most relevant configurations contributing to the total wave function. 
Because of the permutation symmetry, the spatial wave function satisfies 
$\Psi(x_1,x_2)=\Psi(x_2,x_1)$, thus becoming symmetric under reflection with 
respect to the diagonal $x_1=x_2$. Large values of $\Psi(x,x)$, along this 
diagonal, correspond to ionic or doubly occupied configurations. In contrast, 
large density values $\Psi(x,-x)$ along the $x_1=-x_2$ diagonal, corresponds 
to configurations where the electrons are mostly in opposite (left and right) 
half-planes.
 
In the present calculations, the $x$ coordinates of the centers of the left 
and right dots $x_L=-15$ nm, $x_R=15$ nm are held fixed while that of the 
atom, $x_A=x$, varies.
Large values of $\Psi(x_L,x_L)$, $\Psi(x_R,x_R)$ or $\Psi(x_A,x_A)$ entail a 
doubly occupied configuration at the left dot, the right dot or the atom, 
respectively.

On the other hand, a configuration of one electron in the atom and the other 
in a bond (antibond) between the left and right dots, would be represented 
by

\begin{equation}
\Psi(x_1,x_2) = \left[c_L \varphi_L(x_1) \pm c_R \varphi_R(x_1) \right]\varphi_A(x_2)+ (x_1\leftrightarrow x_2), 
\label{eq bond}
\end{equation}

\noindent where the last term represents a term similar to the first one with 
the variables interchanged, and $\varphi_a$ is a wave function centered 
around $x_a$ ($a=L, R, A$). Then, $\Psi(x_1,x_2)$ will have large values 
close to $(x_L,x_A)$ and $(x_R,x_A)$ with the same or opposite sign for a 
bond or antibond, respectively.

\begin{figure}
\includegraphics[scale=0.35]{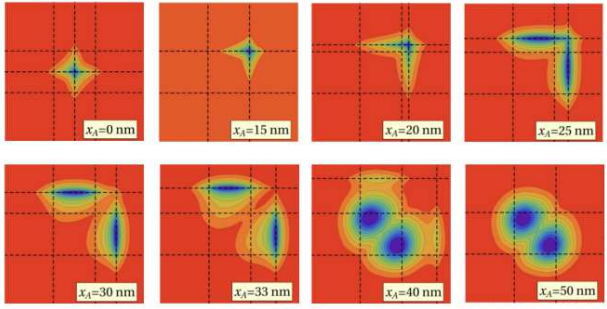}
\caption{\label{2-particle S1} (Color online) Contour plot in the 
$(x_1,x_2)$-plane of the two-electron ground state singlet wave 
function, $\Psi_0^0(x_1,x_2)$, along the interdot axis $x$ for the ground 
state of the doped double quantum dot. 
Gaussian wells are centered at ${\bf R}_L = (x_L,0)$ and ${\bf R}_
R = (x_R,0)$ and the impurity atom of charge $Z=1$, 
at ${\bf R}_A=(x_A,0)$. The vertical and horizontal dashed lines 
$x_i=x_L, x_R$ or $x_A$ ($i =1$, 2), signals the condition where one electron 
(electron 1 or 2, respectively), is at the center of the dot to the left, to 
the right or at the impurity atom.
The centers of the dots $x_R=-x_L=15$ nm are held symmetrical with respect 
to the origin of coordinates. The atom is successively placed at $x_A = 0$, 
$15$, $20$, $25$, $30$, $33$, $40$ and $50$ nm.}
\end{figure}

\begin{figure}
\includegraphics[scale=0.35]{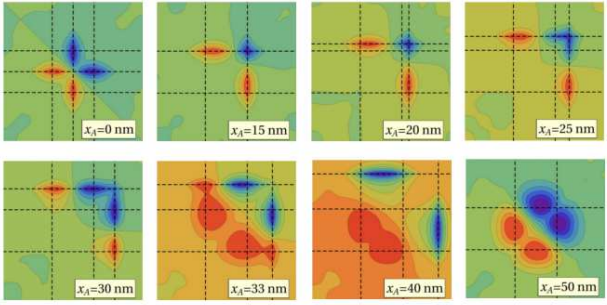}
\caption{\label{2-particle S2} (Color online) Same as Fig. \ref{2-particle S1} 
for the first excited two-electron singlet wave function $\Psi_1^0(x_1,x_2)$.}
\end{figure}

Figs. \ref{2-particle S1} and \ref{2-particle S2} show the plot of the ground 
state $\Psi_0(x_1,x_2)$ and the first singlet excited state 
$\Psi_1(x_1,x_2)$ from our calculations.

For a single-electron symmetric double dot system without impurity, 
the ground and first excited states are the bonding and antibonding states 
formed from the linear combination of orbitals centered at each dot. 
For the two-electron symmetric double dot system, Figs. \ref{2-particle S1} 
and \ref{2-particle S2} show that when the atom is at the center of the 
line joining both dots ($x=0$), the two-particle wave function of the ground 
state (excited state) roughly corresponds to one electron in the atom and the 
other in the bond (antibond) of the double dot system, Eq. (\ref{eq bond}). 
Therefore, the matrix element $\langle \Psi_0|x_1+x_2|\Psi_1\rangle$ roughly 
correspond to the sum of those for the atom and the double dot separately.

At $x=15$ nm, the atom is at the center of the dot to the right, the system 
becomes very asymmetric, with the potential of dot to the right deeper than 
the one to the left due to the contribution of the attractive impurity. The 
bond becomes a doubly occupied state localized close to the center of the 
combined potential (QD$_R$ and impurity), while the antibond becomes more 
localized around to QD$_L$, due the orthogonality condition, what lowers 
$f_{12}$. The behavior  in the range $0\leq x \leq 15$ nm reflects this 
gradual change.

From 15 to 30 nm, the effect of the impurity turns weaker as the atom moves 
away, and the two dots becomes more symmetric again; this redistribute the 
charge towards QD$_L$, recovering some bonding and antibonding character 
for $\Psi_0$ and $\Psi_1$, respectively. Such a configuration favours an 
increase of $f_{12}$. Furthermore, as one electron remains in the atom, which 
is farther from the origin, the matrix element of $x$ becomes larger than the 
one corresponding to the atom at the origin. The oscillator strength has a 
peak at 33 nm roughly increasing quadratically with the position of the atom 
as a consequence of the stretching of the charge.

After 33 nm, the electron in the atom cannot be retained by the impurity 
potential, thus $\Psi_0$ approaches to a configuration with one electron in 
each dot. Nevertheless, the excited state $\Psi_1$ still have a configuration 
where the atom is occupied, what lowers $f_{01}$.

For atom positions further than 40 nm, the excited state also releases its 
electron and the double quantum dot becomes even more symmetric, approaching 
its behavior in the absence of impurity, thus approaching its value 
$f_{01}=2$. The limit of isolated dots is clearly seen in Figs. 
\ref{2-particle S1} and \ref{2-particle S2}, where for $x\gtrsim 50$ nm, 
the ground and first excited states are, approximately,  
$\Psi_0\approx [\varphi_L(x_1)\varphi_R(x_2)+\varphi_R(x_1)\varphi_L(x_2)]/
\sqrt{2}$ and $\Psi_1\approx [\varphi_L(x_1)\varphi_L(x_2)-\varphi_R(x_1)
\varphi_R(x_2)]/\sqrt{2}$.

As seen from Fig. \ref{OS vs x}, the oscillator strength for impurity charges 
smaller than $Z=1$, has simpler features. Basically, they start from a value 
$f_{12}$ slightly less than 2, decreases until a minimum as the atom approaches 
one dot, say QD$_R$, and increases again smoothly until reaching the 
asymptotic impurity-free value of 2.

The oscillator strength is a highly sensitive property to the 
presence of the impurity. A value of $f_{12}$ close to 2, occurs either when 
the impurity is weak wherever it is located, or when a highly charged impurity 
atom is far away from the double QD.  Both cases are situations where the 
impurity is a perturbation for the coupled QDs and, therefore amenable of use 
in quantum computing. On the contrary, deviations of the oscillator strength 
from a value of 2, provides an indication of a breakdown of the possibility to 
consider the system as a double QD.

\section{Conclusions} 

In this work, we have studied the influence of a Coulomb atomic impurity 
on the entanglement entropy of two-dimensional two-electron double quantum 
dots.
The electronic structure was calculated by using a configuration interaction 
method with a Gaussian basis set expansion. 
The degree of entanglement becomes highly modulated by both the location 
and charge screening of the impurity atom. Two regimes are clearly identified: 
one of low entanglement and other of high entanglement, with both of them 
mainly determined by the magnitude of the charge. The exchange coupling 
between the electrons, being proportional to singlet-triplet exchange coupling
, has 
an opposite behaviour with respect to the one of the entropy. 
The efficient use of double QDs with impurities, in specific quantum 
information processing tasks could require the tuning of the interdot 
separation or the quantum well depths, for optimizing the harnessing of the 
entanglement, the exchange coupling  or both.  
Finally, the magnitude of the oscillator strength of the system could provide 
an indication of the presence and characteristics of impurities that could 
largely influence the degree of entanglement of the system. It is clear that
experimentally obtained optical properties can help in the design of double 
QDs with
desirable properties in order to use them in quantum information tasks.

The quantum control of these kind of systems can be implemented using pulses 
of external fields. This issue if of great importance in quantum computation 
and a work about the interaction of electromagnetic fields with the system 
presented here is in progress. Of course the decoherence, not considered in 
these works, plays a very important role in the quantum dynamics of these 
kind of devices. Studies in Markovian scenarios suggested that the 
entanglement vanishes due to decoherence, while the decoherence process in 
non-Markovian regimes sometimes gives rise to an interesting new effect: 
entanglement sudden revival \cite{dec7,nm1,nm2,nm3}. We think that can be 
very interesting and useful to study 
the non-Markovian dynamics of these system in the low and high 
entanglement regions depicted in the present article.

\ack{Acknowledgments}
We would like to acknowledge CONICET (PIP 11220090100654/2010), SGCyT(UNNE) and FONCyT (PICT-2011-0472) 
for partial financial support of this project.


\section*{References}

\begin{thebibliography}{99}


\bibitem{QD1} W. G. van der Wiel, S. De Franceschi, J. M. Elzerman, 
T. Fujisawa, S.Tarucha and L. P. Kouwenhoven, Rev. Mod. Phys.
{\bf 75}, 1 (2003); R. Hanson, L. P. Kouwenhoven, J. R. Petta, 
S. Tarucha and L. M. K. Vandersypen,  Rev. Mod. Phys. {\bf 79}, 1217 (2007);
S. J. Lee, S. Souma , G. Ihm and K. J. Chang, Phys. Rep. {\bf 394}, 1 (2004).


\bibitem{bast} G. Bastard, Phys. Rev. B {\bf 24}, 4714 (1981) 

\bibitem{dasarma1}  Nga T. T. Nguyen and S. Das Sarma, Phys. Rev. B {\bf 83},
235322 (2011)


\bibitem{Ashoori92} R.~C.~Ashoori, H.~L.~Stormer, J.~S.~Weiner, L.~N.~Pfeiffer, S.~J.~Pearton, K.~W.~Baldwin, K.~W.~West, Phys.\ Rev.\ Lett. {\bf 68}, 3088 (1992).

\bibitem{im1} Y. Wan, G. Ortiz, and P. Phillips, Phys. Rev. B {\bf 55},
5313 (1997).

\bibitem{im2} E. Lee, A. Puzder, M. Y. Chou, T. Uzer, and D. Farrelly,
Phys. Rev. B {\bf 57}, 12281 (1998).


\bibitem{ex5} E. R\"as\"anen, J. K\"onemann, R. J. Haug, M. J. Puska, and
R. M. Nieminen, Phys. Rev. B {\bf 70}, 115308 (2004).


\bibitem{mod1} X. Hu and S. Das Sarma, Phys. Rev. A {\bf 61}, 062301 (2000).


\bibitem{ex1} P. A. Sundqvist, V. Narayan, S. Stafstr\"om, and 
M. Willander, Phys. Rev. B {\bf 67}, 165330 (2003)

\bibitem{ex2} S. V. Nistor, M. Stefan, L. C. Nistor, E. Goovaerts, and 
G. Van Tendeloo, Phys. Rev. B {\bf 81}, 035336 (2010)


\bibitem{ex3} V. Narayan and M. Willander, Phys. Rev. B {\bf 65}, 
125330 (2002)

\bibitem{PettaDec} R. Petta, A. C. Johnson, J. M. Taylor, E. A. Laird, 
A. Yacoby, M. D. Lukin, C. M. Marcus, M. P. Hanson, and A. C. Gossard,
Science {\bf 309}, 2180 (2005).

\bibitem{dec1} M. Schlosshauer, Rev. Mod. Phys. {\bf 76}, 1267 (2005).

\bibitem{dec2} C. H. Bennett, Phys. Today. {\bf 48}, 24 (1995). 

\bibitem{dec3} J. Xu, X. Xu, C. Li, C. Zhang, and X. Zou, G. Guo,
Nat. Commun. {\bf 1}, 1 (2010).

\bibitem{dec4} T. Yu, and J. H. Eberly, Phys. Rev. B {\bf 66}, 193306 (2002).  

\bibitem{dec5} A. Ferr\'on, D. Dom\'{\i}nguez, and M. J. S\'anchez, 
Phys. Rev. Lett. {\bf 109}, 237005 (2012).

\bibitem{dec6} G. A. \'Alvarez, and D. Suter, Phys. Rev. Lett. {\bf 104}, 
230403 (2010).

\bibitem{dec7} A. P. Majtey, and A. R. Plastino, Int. J. Quanum Inform. {\bf 
10}, 1250063 (2012)



\bibitem{tichi}M. Tichy, F. Mintert, and A. Buchleinter, J. Phys. B: At. Mol.
Opt. Phys. {\bf 44}, 192001 (2011)

\bibitem{amico} L. Amico, L. Fazio, A. Osterloh, and V. Vedral, Rev. Mod. Phys.
{\bf 80}, 517 (2008)

\bibitem{damico1} S. Abdullah, J. P. Coe and I. D’Amico, Phys. Rev. B {\bf 80},
235302 (2009)

\bibitem{zu1} L. He, and A. Zunger,  Phys. Rev. B {\bf 75}, 075330 (2007)

\bibitem{zu2} L. He, G. Bester, and A. Zunger,  Phys. Rev. B {\bf 72}, 195307 
(2005)

\bibitem{damico2} J. P. Coe, and I. D’Amico, J. Phys.: Conf. Ser. {\bf 254},
012010 (2010)

\bibitem{damico3}J. P. Coe, A. Sudbery, and I. D’Amico, Phys. Rev. B {\bf 77}, 
205122 (2008)

\bibitem{os1}O. Osenda, and P. Serra, Phys. Rev. A {\bf 75}, 042331 (2007).

\bibitem{os2} O. Osenda, and P. Serra, J. Phys. B: At. Mol.
Opt. Phys. {\bf 41}, 065502 (2008).


\bibitem{fos1}A. Ferr\'on, O. Osenda and P. Serra, Phys. Rev. A {\bf 79},
032509 (2009).

\bibitem{pots} F. M. Pont, O. Osenda, J. H. Toloza and P. Serra
Phys. Rev. A {\bf 81}, 042518 (2010).

\bibitem{pos} F. M. Pont, O. Osenda, and P. Serra,
Phys. Scr. {\bf82}, 038104 (2010).

\bibitem{ana} A. P. Majtey, A. R. Plastino, and J. S. Dehesa,
J. Phys A: Math. Theor. {\bf 45}, 115309 (2012).

\bibitem{opt1} J. L. Gondar and F. Comas, Physica B {\bf 322},
413 (2003).

\bibitem{opt2}  S. Yilmaz and H. Safak, Physica E {\bf 36},
40 (2007).

\bibitem{opt3}  A. \"Ozmen, Y. Yakar, B. Cakir and \"U. Atav, Opt. Commun,
{\bf 282}, 3999 (2009).

\bibitem{opt4}  I. Karabulut and S. Baskoutas, J. Appl. Phys {\bf 103},
073512 (2008).

\bibitem{opt5} M. Sahin, Phys. Rev. B {\bf 77}, 045317 (2008).


\bibitem{opt6} J. S. deSousa, J. P. Leburton, V. N. Freire, and E. F. daSilva,
Phys. Rev. B {\bf 72}, 155438 (2005).

\bibitem{ex4} V. Nistor, L.C. Nistor, M. Stefan, C.D. Mateescua, 
R. Birjega, N. Solovieva, and M. Nikl, Superlattices Microstruct., {\bf 46}, 
306 (2009).


\bibitem{var1} A. T. Kruppa and K. Arai, Phys. Rev. A {\bf 59}, 3556 1999.

\bibitem{sr1} D. S. Acosta Coden, S. S. Gomez, and R. H. Romero, 
J. Phys. B: At. Mol. Opt. Phys. {\bf 44}, 035003 (2011).

\bibitem{sr2} S. S. Gomez, and R. H. Romero,  
Cent. Eur. J. Phys. {\bf 7}, 12 (2009)

\bibitem{nm1} B. Bellomo, R. Lo Franco, and G. Compagno, 
Phys. Rev. Lett. {\bf 99}, 160502 (2007).

\bibitem{nm2} T. Yu, and J. H. Eberly, Science {\bf 323}, 598 (2009).

\bibitem{nm3} F. Lastra, S. A. Reyes, and S. Wallentowitz, J. Phys. B: 
At. Mol. Opt. Phys. {\bf 44}, 015504 (2011).







\end{thebibliography}
\end{document}